\documentclass[12pt]{article}

\usepackage{graphicx}

\newcommand\abs[1]{\left|#1\right|}

\newcommand{\beq}{\begin{equation}}
\newcommand{\eeq}{\end{equation}}
\newcommand{\beqa}{\begin{eqnarray}}
\newcommand{\eeqa}{\end{eqnarray}}
\newcommand{\beqar}{\begin{eqnarray*}}
\newcommand{\eeqar}{\end{eqnarray*}}

\begin{document}
\thispagestyle{empty}

\hfill{\sc IFIC/16-31}

\vspace{32pt}
\begin{center}
{\textbf{\Large Deuteron structure in the deep inelastic regime}}

\vspace{40pt}

C.A.~Garc\'\i a Canal$^1$, T.~Tarutina$^1$, V.~Vento$^2$
\vspace{12pt}

\textit{$^1$IFLP/CONICET and Departamento de F{\'\i}sica}\\ 
\textit{Universidad Nacional de La Plata, C.C.67, 1900, 
La Plata, Argentina}\\
\vspace{10pt}
\textit{$^2$Departamento de F{\'\i}sica Te\'orica-IFIC. Universidad de Valencia-CSIC.}\\ 
\textit{E-46100, Burjassot (Valencia), Spain}
\end{center}

\vspace{40pt}

\date{\today}

\begin{abstract}
We study nuclear effects on the deuteron in the deep inelastic regime using the newest available data analyzing their $Q^2$ dependence. We conclude
that precise EMC ratios for large $Q^2$ ($> 30\, $GeV$^2$) cannot be obtained without considering these nuclear effects. For this purpose we use a scheme which parametrizes these effects in a simple manner and compare our results with other recent proposals.

\end{abstract}

\newpage

\section{Introduction} 
The study of nuclear effects in  structure functions are necessary to understand the microscopic structure of nucleons and nuclei in terms of Quantum Chromodynamics \cite{Kuhn:2015lja}.
However, when studying the nucleon structure functions, the  main difficulties are related to the neutron structure function $F_2^n$, because neutrons cannot be prepared as scattering  targets. Consequently, $F_2^n$ has to be extracted from the measurable deuteron structure function $F_2^D$, plus the knowledge of the proton one $F_2^p$.
In doing this analysis one is always facing the problem of quantifying the nuclear structure
effects in the deuteron.  A priori it appears as a reasonable approximation to consider
the deuteron as a free proton plus a free neutron system,  because the binding energy is small 
($2.22$  MeV). Nevertheless, deuteron is not strictly a superposition of free constituents and for this reason the smearing 
produced by nuclear binding effects have been the subject of several analysis based on different physical considerations \cite{previous1,previous2}. These studies ended with a large variety of values for the neutron structure function, all coming from the same experimental data.  In addition to limiting our ability to extract the neutron structure function, the large spread of
results --even among extractions including only traditional nuclear effects such as Fermi motion and binding-- has
made it difficult to identify a reliable baseline which could be used to search for more involved nuclear effects such as
the so called EMC effect \cite{Aub83}.

The new measurements on light nuclei \cite{See09} have generated a renewed interest in the EMC effect even in the polarized case \cite{Cloet:2005rt,Smith:2005ra,Ganesamurthy:2011zza,Fan14}. EMC ratios are usually taken with respect to the deuteron, but the deuteron may also exhibit an EMC effect. Several attempts \cite{previous1,Wei11} have been made to determine $R^d_{EMC}= F^d_2/(F^n_2 +F^p_2)$ where $F^n_2(F^p_2)$ are the free neutron(proton) structure functions. A good experimental determination of $R^d_{EMC}$ can shed some light on the cause of this effect. The high quality data of BONuS \cite{Fen08,Bai12,Tka14} designed to measure $F_2^n/F^p_2$ at high $x$ allows a better determination of $R^d_{EMC}$ \cite{Gri15}. On the other hand new parametrizations of the nucleon structure functions have appeared  \cite{mstw} which allow the study of observables with higher precision and up to higher $Q^2$. This wealth of data has prompted us to review a description presented some time ago \cite{previous1} aimed at making compatible the Gottfried sum rule with the data by considering nuclear effects in the deuteron.  This should allow to identify a reliable baseline which could be used to search for nuclear effects such as
the modification of the nucleon structure function in nuclei or non-nucleonic degrees of freedom.

In the next section we recall the previously mentioned analysis. In section 3 we show the results obtained under this scheme but using all the presently available data and we make a comparison with recent related proposals. We finish by drawing some conclusions.

\section{Gottfried sum rule and nuclear effects in the deuteron}

When the NMC collaboration \cite{NMC} presented in 1991 the analysis of the ratio of
the structure functions $F_2^n/ F_2^p$ obtained in deep inelastic
scattering of muons on hydrogen and deuterium
targets, exposed simultaneously to the beam,
assumed that nuclear effects were not significant in
deuterium, namely
\begin{equation}
F_2^D = \frac{1}{2} \,\left( F_2^p +F_2^n \right)
\label{formula1}
\end{equation}
and consequently 
\begin{equation}
\frac{F_2^n}{F_2^p} = 2\,\frac{F_2^D}{F_2^p} - 1.
\label{formula2}
\end{equation}
The formula (\ref{formula2}) was used in Ref.\cite{NMC} to extract the $F_2^n/F_2^p$ ratios from the experimental data on $F_2^D/F_2^p$.

That data set was also used in order to test the validity of the Gottfried sum rule \cite{Got}
\begin{equation}
\int_0^1\,\frac{dx}{x}\,\left[F_2^p(x) - F_2^n(x)\right] = \frac{1}{3}
\end{equation}
and found a value significantly below this quark-parton model prediction of $1/3$.

Assuming the convergence of the Gottfried sum rule for the extraction
of $F_2^p(x)- F_2^n(x)$ from the data on deuterium \cite{leader}, it was shown, that even though the corrections to the
naive expression (\ref{formula1}) could be small, their effect is highly amplified in this difference. It was concluded that significant tests of the Gottfried sum rule cannot be made on the basis of the deuteron data without considering nuclear effects \cite{leader}.
These measurable nuclear effects in deuterium were in agreement with predictions of several models such as the light cone approach to the deuteron structure function \cite{Kap}, parton recombination model \cite{Close} and pionic effects in the deuteron \cite{Uchi}. On the other hand, the picture that emerged when comparing nuclear structure functions with those of free protons was different from the standard comparison with deuterium protons \cite{aproton}.

In Ref. \cite{previous1} the nuclear effects in deuterium were taken into account  by defining the bound nuclear structure function, $F_2^D$, by means of 
\begin{equation}
F_2^D = \frac{1}{2}\left(F_2^{\prime p} + F_2^{\prime n}\right)\,\,\,\,;\,\,\,\,with\,\,\,\,
F_2^{\prime p} = \frac{1}{\beta} F_2^p.
\end{equation}
Due to isospin symmetry, the $\beta$ factor was taken the same for the  proton  and neutron structure functions.
Then, the difference between the bound nucleon structure functions was expressed as
\begin{eqnarray}  \label{difprime}
(F_2^{\prime p} - F_2^{\prime n} )&=& 2\,F_2^D\,\frac{1 - F_2^{\prime n}/F_2^{\prime p}}
{1 + F_2^{\prime n}/F_2^{\prime p}} 
 =  \frac{1}{\beta}\,\left[\frac{1}{3}\,x\,(u_v - d_v) + \frac{2}{3}\,(\bar{u}  - \bar{d})\right].
\end{eqnarray}
The ratio $ F_2^{\prime n}/F_2^{\prime p}$ is related to experiments by
\begin{eqnarray}
\frac{F_2^{ n}}{F_2^{ p}}\Bigg|_{exp} & = & 2 \frac{F^2_2}{F^p_2} - 1 =   \frac{F_2^{\prime n}}{F_2^{\prime p}} + \frac{1}{\beta} -1.
\end{eqnarray}
Unfortunately, data coming from deuteron targets are always used in the fits although the inclusion of $\beta$ does not alter the conclusions. This equation leads to
\begin{equation}
(F_2^{\prime p} - F_2^{\prime n}) = 2\,F_2^D\,\left(\frac{2}{\beta\,(\frac{F_2^n}{F_2^p}\bigg|_{exp} +1)}
-1 \right).
\label{betacalc}
\end{equation}
From which the parameter $\beta$ can be adjusted by using the experimental data on the deuteron combined with a parametrisation of the quark distributions. 

\begin{figure}[t]
\centerline{\includegraphics[scale=0.4]{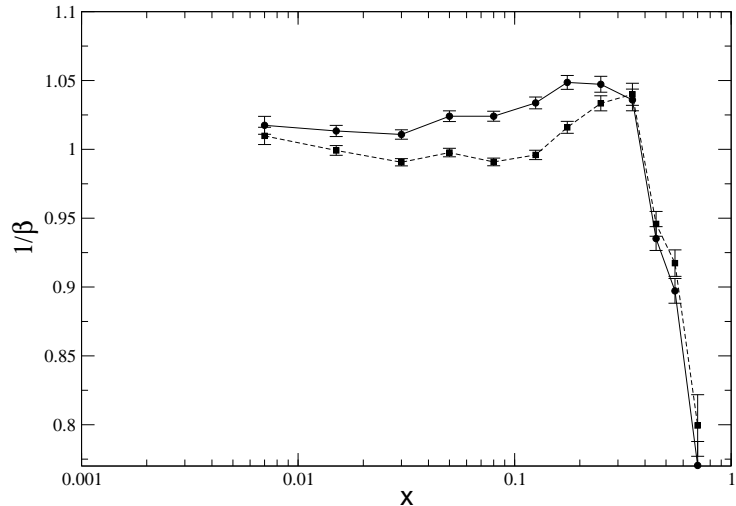}}
\vskip 1cm
\caption{The values of $1/\beta$ calculated using NMC \cite{NMC} data and Morfin and Tung \cite{morfin} (filled circles) and MSTW \cite{mstw} (filled squares) distributions. The lines are shown to guide the eye.}
\label{fig1}
\end{figure}

In Fig.\ref{fig1} we show the original calculation of the parameter $1/\beta$ as a function of Bjorken $x$ from Ref.\cite{leader} using  Eq.(\ref{betacalc}) and one using a new parametrisation of the structure functions. We use  Eq.(\ref{formula2}) with the experimental data on the ratio $F_2^n/F_2^p|_{exp}$  from  Ref.\cite{NMC} and the absolute deuteron structure function from a fit to published data from other experiments. For quark distributions we show two parametrizations: the Morfin and Tung parametrization ($s$-fit in the DIS scheme) \cite{morfin}, shown with  filled circles and the more recent MSTW parametrization \cite{mstw} shown with filled black squares . The lines are to guide the eye. See Ref. \cite{arrington} for a clear explanation on the downturn of the low $Q^2$ data associated to target mass corrections.  Some features of nuclear effects are apparent in Fig. \ref{fig1}. The antishadowing maximum appears clearly in both parametrizations around $x=0.2$ persisting for low x with Morfin and Tung \cite{morfin},  but not so with MSTW \cite{mstw}, where it disappears for $x<0.15$. 

As it was mentioned, a remarkable feature of the Gottfried sum rule is that it is an amplifier of nuclear effects.  A small amount of antishadowing  causes a big deviation  in the integrand. This explains why $\beta$ is quite independent of the parton distribution used and why the deuteron nuclear effects could be safely neglected in many analysis. 

The above results  show that structure function ratios with respect to deuterium is not a precise procedure to extract nuclear effects  since the composite nature of deuterium at the nuclear level has to enter into the theoretical description for those effects. It is therefore relevant to see how the new data complete this picture.

\section{Revisiting nuclear effects in the deuteron}

The study of nuclear effects in the deuteron has been a subject which has gained great interest in the last years. Other schemes have completed the description above and we proceed to discuss them next and to compare them with our scheme using the latest available data.

In Ref.\cite{martin} nuclear corrections were defined in terms of a function $c(x)$ by means of
\begin{equation}
F^D(x,Q^2)=c(x)\left[F^n(x,Q^2)+F^p(x,Q^2)\right]/2
\end{equation}
from this definition it is clear that this correction $c(x)$ is, in principle, essentially equal to our $1/\beta$ factor.

In Ref.\cite{martin} the following parametrisation of $c(x)$ was presented
\begin{equation}
c(x) = \left\{ \begin{array}{ll}
 (1+0.01N)(1+0.01c_1\ln^2(x_p/x)) &\mbox{ if $x<x_p$} \\
 (1+0.01N)(1+0.01c_2\ln^2(x/x_p)+0.01c_3\ln^{20}(x/x_p)) &\mbox{ if $x>x_p$}
       \end{array} \right.
 \label{martin}
\end{equation}
where $x_p$ is a 'pivot point' at which the normalization is $(1+0.01N)$. The values of the parameters for the deuteron correction factor 
are given in Table 1 of Ref.\cite{Har14}.

In Ref.\cite{Gri15}  the structure function ratio $R_{EMC}^d=F_2^d/(F_2^n+F_2^p)$ was computed and can easily be connected with our $\beta$,
\begin{equation}
R_{EMC}^d=F_2^d/(F_2^n+F_2^p)=2F_2^D/(F_2^n+F_2^p)=(F_2^{\prime p} + F_2^{\prime n})/(F_2^n+F_2^p)=1/\beta.
\end{equation}
The deuteron structure function $F_2^D$ defined in Eq.(\ref{formula1}) is a structure function per nucleon and it is connected with the $F_2^d$ used in Ref.\cite{Gri15} by
\begin{equation}
F_2^d=2F_2^D.
\end{equation}
These authors have used the recently published data on $F_2^n/F_2^d$ taken by the BONuS experiment using CLAS at Jefferson Lab \cite{Fen08,Bai12,Tka14}. For $F_2^p/F_2^d$ they used the available global parametrisations.

In Ref. \cite{Wei11} they describe the so called IMC (In-Medium Correction) effect in terms of the slope of the EMC effect arriving to the conclusion that 
\begin{equation}
\abs{\frac{d R_{IMC}(A)}{dx} } = \abs{ \frac{d R_{exp}(A)}{dx} }  +0.079 \pm 0.006.
\end{equation}
Their main result in this respect is their Fig.2.

\begin{figure}[t]
\centerline{\includegraphics[scale= 0.6]{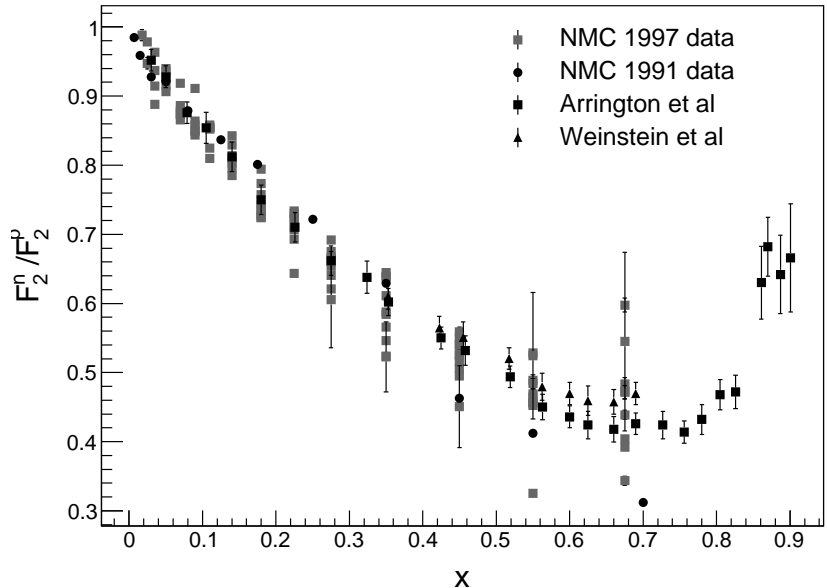}}
\vskip 1cm 
\caption{Ratios $F_2^n/F_2^p$ obtained from experimental data of Ref.\cite{arneodo} presented on previous figure (shown here with grey filled squares for all values of $Q^2$). We also present the older data of NMC \cite{NMC} (shown with black filled circles) together with (a) the ratios of $F_2^n/F_2^p$ obtained by Arrington et al. from various experiments in Ref.\cite{arrington} (shown with black filled squares); (b) the ratios using the data of et Weinstein al. \cite{Wei11} (shown with black filled triangles). }
\label{fig2}
\end{figure}

In Fig.\ref{fig2} we show the ratios $F_2^n/F_2^p$ obtained from the experimental ratios $F_2^D/F_2^p$ of Ref.\cite{arneodo} using Eq.(\ref{formula2}). We also show the older NMC data given in Ref.\cite{NMC} used in the previous estimation of $\beta$. In this Figure we also present (a) the ratio  $F_2^n/F_2^p$ extracted in Ref.\cite{arrington} that makes use of data available from different experiments and
(b) the ratio using the data of Weinstein et al. \cite{Wei11}.  The figure shows the consistency between all the modern data and makes apparent that the old data, available only in the region $x<0.7$, give a value for the ratio smaller that that from the modern data. 

We begin by presenting the results in our $\beta$ scheme using the newest available data to proceed thereafter to compare with all the other schemes.


\begin{figure}[t]
\centerline{\includegraphics[scale= 0.6]{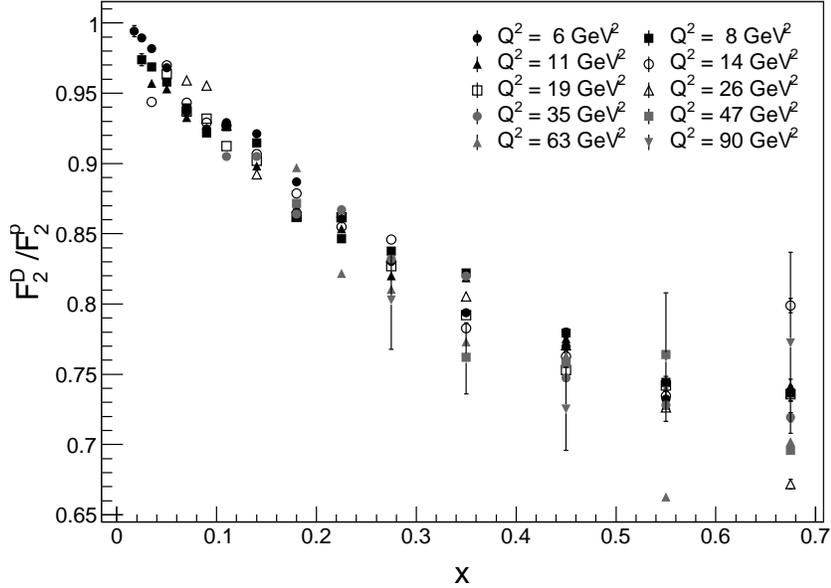}}
\vskip 1cm
\caption{The experimental ratios $F_2^D/F_2^p$ as a function of $x$ obtained by New Muon Collaboration in Ref.\cite{arneodo}. We show explicitly the $Q^2$ dependence of the data.}
\label{fig3}
\end{figure}


We show in Fig.\ref{fig3} the experimental ratios $F_2^D/F_2^p$  obtained by the New Muon Collaboration (NMC) in 1997 and presented in 
Ref.\cite{arneodo}. To study the importance of the $Q^2$ dependence we show the data separately for each value of $Q^2$.


\begin{figure}[t]
\centerline{\includegraphics[scale= 0.6]{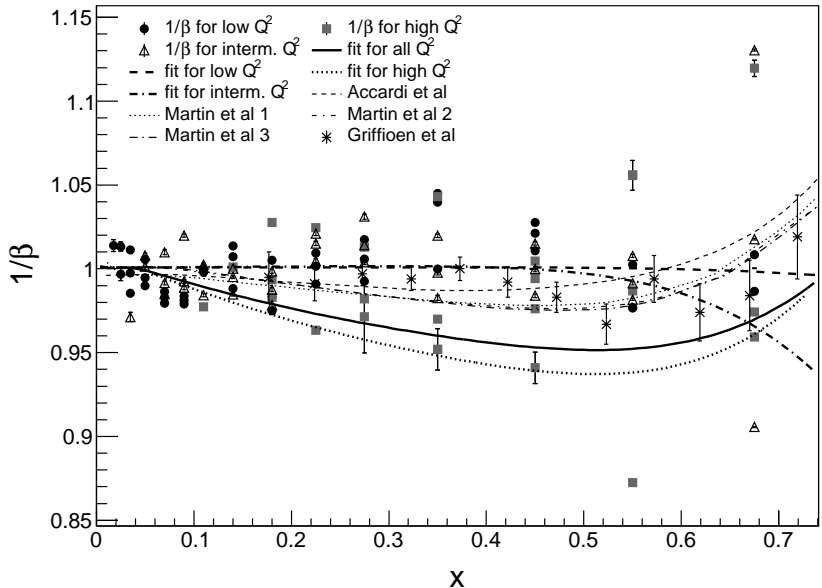}}
\vskip 1cm
\caption{The values of $1/\beta$ obtained from the NMC 1997 data \cite{arneodo} together with the fits and compared to the nuclear corrections calculated in the other works. See text for details.}
\label{fig4}
\end{figure}


From the data in Fig. \ref{fig3} we extract $1/\beta$ using Eq.\ref{betacalc} using the new MSTW parametrization of the structure functions \cite{mstw}. We adjust these data to a functional fit of the type used in Ref. \cite{martin} shown in Eq. (\ref{martin}). We perform four types of fits: 
for low $Q^2 (6, 8, 11$ GeV$^2)$, intermediate  $Q^2$ (14, 19, 26 GeV$^2)$, high $Q^2 (35, 47, 63, 90$ GeV$^2)$ and all $Q^2$. 

Fig.\ref{fig4} summarizes the different nuclear corrections proposed including  our new results for $1/\beta$. The values of $1/\beta$ corresponding to the low $Q^2$ are shown with
black filled circles, the high $Q^2$ values are presented by grey circles and the intermediate  $Q^2$ values are given by the open triangles.
The results of the fit are given by the thick continuous lines: the solid line represents the fit for entire range of $Q^2$, the dashed line -- the fit for low $Q^2$, the dotted line -- the high $Q^2$ and the dash-dotted line stands for the fit for the intermediate region of $Q^2$.
The thin long dashed line is the nuclear correction calculated by Accardi et al in Refs.\cite{Acc16a,Acc16b} for the $Q^2$ = 10 GeV$^2$.
The corrections $c(x)$ calculated for the three sets of parameters given in Ref.\cite{Har14} are presented with continuous lines (solid, dashed and dot-dashed for three different parameter sets). The values of $R_{EMC}^d$ obtained in Ref.\cite{Gri15} are shown with stars and contain error bars.  Our fit for all data is in relatively good agreement with all other results. However, our study of the $Q^2$ dependence indicates that for low $Q^2$ nuclear effects are inexistent, for intermediate $Q^2$ they appear, if at all, at higher $x$ and for high $Q^2$ they are very important.

\section{Conclusions}

We have studied nuclear effects below the Fermi motion dominated region ($x<0.75$) on the deuteron using the newest available data by revisiting a scheme proposed initially when the first NMC data appeared \cite{previous1}. We have compared our $\beta$ scheme with all other analysis appeared recently in the literature and have shown that they agree in general terms. The structure of the deuteron is non trivial and manifests itself quite dramatically in DIS. We have studied the $Q^2$ dependence of its structure effects finding that for low $Q^2$ they are irrelevant in all the studied region, for intermediate $Q^2$ they are irrelevant for most of the studied region ($x<0.5$) and above better data are required to find the precise behavior although one would not expect more than 1\% effect. For high $Q^2$ the effect is 
considerable. Thus our conclusion is that one cannot neglect the nuclear structure of the deuteron when performing EMC ratios at high $Q^2$ ($Q^2 > 30$ GeV$^2)$. The deuteron structure modifies the ratios specially around the antishadowing region and this might impede, if not taken into account, a correct physical interpretation. Thus the nature of the deuteron has to enter the description of the data from any QCD based analysis.

\section*{Acknowledgments}
We would like to thank R. Sassot for illuminating discussions.
The work  has been partially supported by ANPCyT and CONICET of Argentina, and by MINECO (Spain) Grant. No. FPA2013-47443-C2-1-P, GVA-PROMETEOII/2014/066 and  SEV-2014-0398.

\end{document}